\documentclass[aps,preprint,preprintnumbers,amsmath,amssymb]{revtex4}
\usepackage{mathrsfs}
\usepackage{amssymb}
\usepackage{graphicx}% Include figure files
\usepackage{dcolumn}% Align table columns on decimal point
\usepackage{bm}% bold math
\usepackage{color}
\usepackage[breaklinks=true,colorlinks=true,linkcolor=blue,urlcolor=blue,citecolor=blue]{hyperref}
\usepackage{soul}
\usepackage{ulem}
\UseRawInputEncoding   % This is to fix the problem of UTF-8
\makeatletter

\newcommand{\Rmnum}[1]{\expandafter\@slowromancap\romannumeral #1@}

\begin{document}

\title{Pressure-tuning domain-wall chirality in noncentrosymmetric magnetic Weyl semimetal CeAlGe}

\author{Xiaobo He$^{1}$}
\author{Yuke Li$^{2}$}
\author{Hai Zeng$^{1}$}
\author{Zengwei Zhu$^{1}$}
\author{Shiyong Tan$^{3}$}
\author{Yongjun Zhang$^{4}$}
\author{Chao Cao$^{5}$}
\email[]{ccao@zju.edu.cn}
\author{Yongkang Luo$^{1}$}
\email[]{mpzslyk@gmail.com}
\affiliation{$^1$ Wuhan National High Magnetic Field Center and School of Physics, Huazhong University of Science and Technology, Wuhan 430074, China;}
\affiliation{$^2$ School of Physics and Hangzhou Key Laboratory of Quantum Matter, Hangzhou Normal University, Hangzhou 311121, China;}
\affiliation{$^3$ Science and Technology on Surface Physics and Chemistry Laboratory, Mianyang 621908, China;}
\affiliation{$^4$ Institute for Advanced Materials, Hubei Normal University, Huangshi 435002, China;}
\affiliation{$^5$ Center for Correlated Matter and School of Physics, Zhejiang University, Hangzhou 310058, China.}

\date{\today}

\maketitle
{\noindent}	 \rule[-10pt]{16.3 cm}{0.05em}\\

\noindent
Topological magnetic Weyl semimetals have been proposed to host controllable chiral domain walls which bear a great prospect in device applications. To exploit them in applications, it is important to have a proper way to tune and manipulate these domain walls. One possible means is through magnetoelastic coupling. The involvement of rare earth in the lately proposed $R$Al$X$ ($R$=rare earth, $X$=Si and Ge) family magnetic Weyl semimetals may provide such a platform. Here we present transport and thermodynamic properties of CeAlGe under hydrostatic pressure. We find that pressure enhances the antiferromagnetic exchange in CeAlGe but essentially retains its magnetic structure. Large topological Hall effect with pronounced loop shape is observed within the magnetically ordered state, and it splits into two regions under pressure. Such an unusual electromagnetic response is inferred to be a consequence of chiral magnetic domain walls. The unprecedented concomitance of its evolution under pressure and the reentrance of antiferromagnetic order strongly suggest the capability of switching on/off this electromagnetic response in noncentrosymmetric magnetic Weyl semimetals via magnetoelastic coupling.\\

\noindent
\textbf{Magnetic Weyl semimetals, Loop-shaped topological Hall effect, Domain wall, Weyl point, Magnetoelastic coupling}

\noindent
\textbf{PACS number(s):} 71.20.Gj, 75.60.Ch, 72.20.My, 71.20.Eh, 71.15.Mb

%72.20.My   Hall effect (in semiconductors)
%71.15.Mb   Local-density approximation (condensed matter)
%71.20.Eh   Rare earth metals and alloys (electronic structure)
%75.60.Ch   Domain walls and domain structure
%71.20.Gj   Semimetals (electronic structure)

{\noindent}	 \rule[-10pt]{16.3cm}{0.05em}\\
\textbf{Citation:} X. He, Y. Li, H. Zeng, Z. Zhu, S. Tan, Y. Zhang, C. Cao, and Y. Luo, Pressure-tuning domain-wall chirality in noncentrosymmetric
magnetic Weyl semimetal CeAlGe, Sci. China-Phys. Mech. Astron. \textbf{66}, 237011 (2023), https://doi.org/10.1007/s11433-022-2051-4

{\noindent}	 \rule[10pt]{16.3cm}{0.05em}\\

\noindent
\textbf{1 Introduction}\\

\noindent
A magnetic domain wall (DW) is a gradual reorientation of individual magnetic moments across a finite-thickness interface that separates domains of different magnetizations. Such a topological defect in real space is usually a consequence of the interplay among exchange interactions, dipolar interaction and the magnetocrystalline anisotropy\cite{Blundell-MaginCM}. The efficient manipulation of DWs has manifested strong technological interest and bears great potential applications in many current and proposed devices \textit{e.g.} high-density data storage, memristors, tunable radio-frequency signal sources and detectors, etc\cite{Maradudin-SurfSci}.

Recently, theorists have proposed nontrivial controllable DWs present in magnetic Weyl semimetals (WSMs)\cite{Huang-WSMDomain}. WSMs are a class of topological materials whose low-energy electronic dispersion is described by the three-dimensional Weyl Hamiltonian\cite{Weyl1929,Armitage-RMP2018}. Such an electronic state can be realized by breaking either crystalline-inversion\cite{Weng-TmPn,XuS-TaAsARPES,Lv-TaAsPRX,Shekhar-NbP,Nirmal-NbAs,Zhang-TaAsLMR,LuoY-NbAsSdH} or time-reversal symmetry\cite{Xu-HgCr2Se4,Liu-SrMnSb2,Liu-Co3Sn2S2,Zeng-Co3Sn2S2,Weng-Co3Sn2S2,Wang-Co3Sn2S2,Yang-Co3Sn2S2,Kubler_Mn3X,Belopolski-Co2MnGa}, and the latter are usually termed as magnetic WSMs. Amongst the known magnetic WSM candidates, the $R$Al$X$ ($R$=Rare earth, $X$=Si, Ge) family compounds are of special peculiarity, because here crystalline-inversion and time-reversal symmetries are simultaneously broken\cite{ChangG-RAlGeDFT}. The magnetic exchange further stabilizes the Weyl nodes that have been already generated by the noncentrosymmetric $I4_1md$ (No. 109) crystalline structure. Indeed, the topological features of $R$Al$X$ compounds were demonstrated by photoemission\cite{Xu-LaAlGe,Sanchez-PrAlGeARPES}, quantum oscillations\cite{Su-LaAlSiSdH,Yang-CeAlSiLHE,Piva-CeAlSiPressure,Lyu-PrAlSi,Wu-PrAlSiLT} and emergent electromagnetic responses such as topological Hall effect (THE)\cite{Puphal-CeAlGeTHE} and anomalous Hall effect (AHE)\cite{Yang-CeAlSiLHE,Piva-CeAlSiPressure,Lyu-PrAlSi,Yang-PrAlGe_Si}. In addition, the inclusion of rare-earth elements endows them with strong magnetoelastic coupling and thus high tunability, providing an ideal platform to investigate the pressure-controlled DW landscapes. Although recent singular angular magnetoresistance\cite{Suzuki-CeAlGeADMR}, scanning superconducting quantum interference device (SQUID) microscope\cite{Xu-CeAlSiDW} and vector magneto-optical Kerr effect (MOKE)\cite{Sun-CeAlSiDW} measurements have confirmed the existence of magnetic DWs in CeAlGe and CeAlSi, the effect of pressure on these DWs has been seldom studied\cite{Piva-CeAlSiPressure}.

Here we systematically study the transport and thermodynamic properties of CeAlGe under hydrostatic pressure up to 2.52 GPa. We find that pressure enhances the antiferromagnetic (AF) transition temperature but essentially retains its magnetic structure, similar to the pressure effect in CeAlSi\cite{Piva-CeAlSiPressure}. More interestingly, a clear loop-shaped topological Hall effect (LTHE, which is a consequence of chiral DWs) observed between 0.15 and 0.7 T at ambient pressure is gradually split and well separated into two regions under pressure. We argue that such an evolution of LTHE with pressure is related to the reentrance of AF phase out of a field-induced feromagnetic (FIF) phase under high pressure. Our results strongly suggest the tunablity of domain wall chirality with the application of pressure, and manifest the great prospects of magnetic WSMs in device application.\\

\noindent
\textbf{2 Methods}\\

\noindent
Single crystals of CeAlGe were grown by flux in the molar ratio 1:1.1:20 placed in alumina crucibles under ultrahigh-vacuum-compatible conditions. The mixture was heated to 1448 K in 10 h, held for 6 h and then slowly cooled at 1 K/h down to 973 K, at which the Al flux was effectively removed by centrifugation. The obtained crystals were in a plate shape with a typical dimension of 4 $\times$ 4 $\times$ 1 mm$^3$. Further, the energy-dispersive X-ray spectroscopy (EDS) confirms the stoichiometry of Ce, Al and Ge as 1.04(1), 1.06(1) and 0.90(1), respectively.

The magnetic transition and isothermal field dependent magnetization of CeAlGe at ambient were measured in a commercial magnetic property measurement system (MPMS-VSM, Quantum Design). Two samples - labeled as S1 and S2 - were screened at ambient pressure by transport measurements, while pressure experiments were performed only on S1. A hybrid piston-clamp type cell was employed to generate hydrostatic pressures up to 2.52 GPa,  with Daphne oil 7373 used as the pressure medium, and pressure in the cell was determined from the SC transition of Pb. Heat capacity under pressure was measured by an AC calorimetric method, in which Chromel-Au$_{99.93\%}$Fe$_{0.07\%}$ thermocouple was used to measure the heat-temperature response. For electric transport measurements under pressure, ohmic contacts were prepared by spot-welding in a Hall-bar geometry, and both in-plane electric resistivity ($\rho_{xx}$) and Hall resistivity ($\rho_{yx}$) were measured in a lock-in amplifier (SR865A equipped with SR554A pre-ampfilier). For all these measurements, the magnetic field was applied along [001].

The lattice constants and internal coordinates of CeAlGe at different pressures were fully optimized using density functional theory code Vienna Abinit Simulation Package (VASP) until total forces on each atom smaller than 1 meV/\AA\ and internal stress less than 0.1 kBar\cite{method:vasp,method:pawvasp}. During the optimization, the Ce-4$f$ orbitals were treated as core-states. The energy cutoff of plane-wave basis were chosen to be 480 eV, and a $\Gamma$-centered $9\times9\times9$ K-mesh was used for the Brillouin Zone integration. The optimized structure was then employed in the magnetic state calculations using full-potential linearized augmented plane-wave method as implemented in Elk code\footnotemark[1]\footnotetext[1]{http://elk.sourceforge.net/.}. In the magnetic state calculations, the DFT+$U$ method was applied on Ce-4$f$ states with $U_f=$6.0 eV and $J_f=$0.7 eV and fully localized limit (FLL) double counting scheme\cite{method:flldc}. The $R(G+k)_{\mathrm{max}}$ was set to 9 with largest muffin-tin radii of 2.5 Bohr for Ce. The Perdew, Burke and Ernzerhoff flavor of generalized gradient approximation to the exchange-correlation functional\cite{method:pbe} was employed. The FP-LAPW results were also compared with VASP calculations using DFT+$U$ method with same parameters. The SOC was considered in all calculations using a second variational method. Using the maximally projected Wannier function method\cite{method:wannier90,method:mpwf}, we obtained effective Hamiltonian using Ce-5$d$, Ce-4$f$, Al-2$p$, and Ge-4$p$ orbitals. The resulting Hamiltonian were then symmetrized using full crystal and magnetic symmetries\cite{ZHI2022108196}, which was then applied to the following calculations of topological properties\cite{WU2017}.\\

\noindent
\textbf{3 Results and discussion}\\

\noindent
The CeAlGe crystals we studied were grown by an Al self-flux method as described in \textbf{Method}. Energy-dispersive X-ray spectroscopy (EDS) confirms the atomic ratio of Ce:Al:Ge = 1.04(1):1.06(1):0.90(1) [Fig.~S1(a) in \textbf{Supplementary Information} (\textbf{SI})], which is among the closest to stoichiometric CeAlGe for grown by Al flux\cite{Hodovanets-CeAlGeSX,Puphal-RAlGe}, so far as we know. The crystalline structure of CeAlGe is displayed in Fig.~1(a). Theory predicted that it is a type-\Rmnum{2} WSM\cite{ChangG-RAlGeDFT}. The magnetic susceptibility of our sample shows a peak at $T_N$=5.1 K due to the formation of long-range AF ordering of the Ce sublattice [cf Fig.~S1(c)], consistent with previous reports\cite{Puphal-RAlGe,Hodovanets-CeAlGeSX}. Neutron scattering experiments by Suzuki \textit{et al} suggested a coplanar but noncollinear magnetic structure with $\mathbf{m_A}=(m_x,m_y,0)$ on sublattice A and $\mathbf{m_B}=(-m_y,-m_x,0)$ on sublattice B in $Fd'd2'$ representation\cite{Suzuki-CeAlGeADMR}, seeing also in Fig.~1(a).

Figure 1(b) and (c) show the temperature dependence of in-plane resistivity ($\rho_{xx}$, sample S1) and AC heat capacity ($C_{ac}$) measured at various pressures. For better clarity, we vertically shifted the curves, and the $\rho_{xx}$ data have been nomalized to the values at 20 K. At ambient pressure, $\rho_{xx}(T)$ decreases monotonically upon cooling until $T\sim$ 5.1 K where $\rho_{xx}(T)$ turns up due to the AF transition\cite{Hodovanets-CeAlGeSX,Puphal-RAlGe}, seeing Fig.~S1(b) in SI. The hump around 120 K in $\rho_{xx}(T)$ arises from the crystalline electric field (CEF) effect as is seen in other cerium-based Kondo lattice compounds. Under pressure, one clearly sees that the AF transition moves to higher temperatures, at a rate of $\sim$0.64 K/GPa [cf Fig.~1(b)]. The same trend is also seen in $C_{ac}(T)$ shown in Fig.~1(c). Such a pressure dependence of $T_N$ is similar as in CeAlSi\cite{Piva-CeAlSiPressure} that can be considered as the chemical pressure effect of CeAlGe, but is relatively unusual \cite{Park-CeRhIn5Pre, Wang-CePd2Al8} for most cerium-based Kondo lattices where pressure typically suppresses magnetic ordering and leads to quantum phase transitions or quantum critical points\cite{Misra-HFSystem}. A possible explanation might be due to the very low carrier density ($n$= 0.063 hole/f.u. from Hall effect, see below) that severely weakens the Kondo coupling (refer to Nozi\`{e}res exhaustion idea or protracted Kondo screening\cite{Nozieres-EPJB1998,LuoY-CeNi2As2Pre}), and therefore, the strength of magnetic exchange increases faster than the Kondo effect under pressure\cite{Doniach}. Another important feature in $C_{ac}$ is that after properly subtracting a $T^3$ term for the phonon contribution, the derived electronic contribution $C_{el}$ can be well scaled in the plot of $C_{el}/T$ vs. $T/T_N$, seeing Fig.~1(d). This manifests that the form of magnetic ordering in CeAlGe remains essentially unchanged with pressure. Note that a similar conclusion was also found in CeAlSi by AC susceptibility measurements\cite{Piva-CeAlSiPressure}.

The magnetic field ($\mu_0H$, refereed to as $H$ henceforth) dependence of Hall resistivity $\rho_{yx}(H)$ (after anti-symmeterized about $H$=0) taken at $p$ = 0 is displayed in Fig.~2. Most critically, a pronounced hysteresis loop can be observed in $\rho_{yx}(H)$ between 0.15 and 0.7 T at 2 K, i.e. the $\rho_{yx}(H)$ curve for field sweeping from positive to negative does not overlap with that from negative to positive [Fig.~2(a)]. Apparently, such a loop-shaped Hall effect is of magnetic origin, because it fades out as $T$ increases and disappears when above $T_N$, seeing Fig.~2(b). Similar phenomenon was also observed in another crystal S2, while the size and the shape of the loop are sample-dependent [Fig.~S2(b)]. However, we should also emphasize that the $c$-axis magnetization ($M_c$) as a function of $H$ is featureless in this region, implying that the observed loop-shaped signal is not attributable to a standard AHE that generally scales with magnetization\cite{Nagaosa-AHERMP}, but should be categorized as THE. Aside from this LTHE, the $\rho_{yx}(H)$ curve is almost linear in $H$ up to 7 T, whereas $M_c(H)$ tends to saturates above 4 T, meaning that in any case one cannot absorb AHE as one of the contributions to the total Hall signal. The absence of AHE is in contrast to a previous report on the CeAlGe grown by floating zone method\cite{Puphal-CeAlGeTHE} in which $\rho_{yx}(H)$ contains both THE (but not loop-shaped) and AHE in addition to the normal Hall effect\cite{Puphal-CeAlGeTHE}. The reason for such bifurcation is unclear, however, according to P. Puphal's elemental analysis, the floating-zone crystals are much closer to the intended 1:1:1 stoichiometry than the flux-grown crystals. This might affect the scattering mechanism. This is possible, because the residual resistance ratio of floating-zone crystal is $RRR$$\equiv$$\rho_{xx}(300K)/\rho_{xx}(2K)$=1.3 \cite{Puphal-RAlGe}, even smaller than in our sample  [$RRR$=1.85, Fig.~S1(b)], indicating that the scattering rate there is higher. Since the strength of AHE $\rho_{yx}^{A}\propto\rho_{xx}^{\alpha}$ ($\alpha$=1 for skew scattering, and $\alpha$=2 for intrinsic and side-jump)\cite{LeeW-CuCr2Se4AHE,Nagaosa-AHERMP}, qualitatively it is reasonable to expect that the AHE in our case would be much weaker. It should be mentioned also that in the sister compound CeAlSi, AHE is seen only for $\mathbf{H}\parallel\mathbf{a}$ but not for $\mathbf{H}\parallel\mathbf{c}$.

Apart from this LTHE, $\rho_{yx}(H)$ remains essentially the same with constant slope between 2 K ($< T_N$) and 8 K ($>T_N$), in spite of that the magnetic susceptibility $\chi_c$ changes by a factor of $\sim 70\%$ [Fig. S1(c)], and this again demonstrates that AHE has negligible contribution to the $\mathbf{H}\parallel\mathbf{c}$ Hall resistivity in CeAlGe. The absence of AHE enables us to estimate the carrier density from the normal Hall effect, $n\approx1.00\times10^{21}$ cm$^{-3}$ for S1 and $0.88\times10^{21}$ cm$^{-3}$ for S2, or equivalently $\sim$0.063 hole/f.u. The low carrier density confirms its semimetallic nature, and weakens the electronic correlation effect as discussed above. By subtracting the normal Hall effect, we obtain the LTHE $\rho_{yx}^{T}$ as a function of field as shown in the bottom panel of Fig.~3(c). For additivity, we convert the Hall resistivity into Hall conductivity via $\sigma_{xy}=\rho_{yx}/(\rho_{xx}^2+\rho_{yx}^2)\approx\rho_{yx}/\rho_{xx}^2$, and the results are depicted in Fig.~2(c). $\sigma_{xy}$ can be decomposed into two contributions, the normal Hall conductivity ($\sigma_{xy}^{N}$) and the topological Hall conductivity ($\sigma_{xy}^{T}$). This analysis yields the magnitude of $\sigma_{xy}^T$ as large as $\sim$110 $\Omega^{-1}\cdot$cm$^{-1}$, exceeding 10\% of the $\nu=1$ quantum Hall effect ($e^2/h$, where is $e$ is elementary charge, and $h$ is Planck's constant) per atomic layer.

We should point out that a similar LTHE was firstly observed in a pyrochlore-type AF insulator Nd$_2$Ir$_2$O$_7$\cite{Disseler-Nd2Ir2O7LHE,Ueda-Nd2Ir2O7LHE}. Earlier theories have predicted the possibility of WSM state in this compound when electronically tuned (\textit{e.g.} pressure) \cite{Wan-IrTSM,Witczak-IrPre,Piva-CeAlSiPressure}. The DWs at the boundary between all-in-all-out domains pin in-gap Weyl fermions and the projected Fermi-arc surface states which give rise to an emergent metallic interface\cite{Yamaji-MetallicDW} and a loop-shaped electromagnetic response\cite{Ma-Nd2Ir2O7}. Recently, similar LTHE was also reported in CeAlSi\cite{Yang-CeAlSiLHE,Piva-CeAlSiPressure}, a type-\Rmnum{1} magnetic WSM isostructural to CeAlGe, but there the ground magnetic structure is an in-plane noncollinear ferromagnetic order below 8.2 K\cite{Yang-CeAlSiLHE}. In CeAlSi, picoscale-magnetoelasticity governing heterogeneous magnetic domains have been observed by B. Xu \textit{et al} using scanning SQUID microscope\cite{Xu-CeAlSiDW}. Vector MOKE measurements further confirmed two kinds of DWs with distinct topology in CeAlSi\cite{Sun-CeAlSiDW}: a non-chiral DW that form parallel to [110] and a chiral DW alined along the [100] axes. The interaction between Weyl nodes and chiral DW are expected to result in the observed LTHE. Compared with CeAlSi\cite{Yang-CeAlSiLHE,Piva-CeAlSiPressure}, the magnitude of LTHE in CeAlGe reaches 1.1 $\mu\Omega\cdot$cm at 2 K, about three times larger than in CeAlSi.

The common feature in the Hall effects in Nd$_2$Ir$_2$O$_7$, CeAlSi and CeAlGe inspires us the important role of DWs to the LTHE observed in CeAlGe. This idea is preliminary testified by the sample dependence [see Fig. S2(a) in \textbf{SI}], and can be further supported by the measurements under pressure, as pressure possibly modifies the landscape of DWs via magnetoelastic coupling and will change the LTHE correspondingly. A qualitative, simplified understanding of this motivation is through Landau-Ginzburg description, where the total free energy ($F_{tot}$) contains the terms of elastic ($F_{\varepsilon}$), order-parameter ($F_{\mathbf{m}}$) and strain-order-parameter ($F_{\mathbf{m}-\varepsilon}$) energies,
\begin{equation}
\begin{aligned}
    F_{tot}&=F_{\varepsilon}+F_{\mathbf{m}}+F_{\mathbf{m}-\varepsilon}\\
           &=F_0+\frac{1}{2}c_{\Gamma}\varepsilon_{\Gamma}^2+   \frac{1}{2}a(T)\mathbf{m}^2+\frac{1}{4}b\mathbf{m}^4+g_{\Gamma}\varepsilon_{\Gamma}\mathbf{m}_{\Gamma}+...
\end{aligned}
\label{Eq1}
\end{equation}
in which $c$ is elastic constant, $\varepsilon$ is strain, $\mathbf{m}$ is order parameter (meaning sublattice magnetization here), $g$ is magneto-elastic coupling constant, the subscript $\Gamma$ is the irreducible representation considered, $F_0$, $a(T)$ and $b$ are as conventionally defined in Landau-Ginzburg's framework\cite{Luthi-Acoustic}. It is generally believed that $g$ is substantial for rare earth ions due to the small energy scales of $f$ electrons, and this yields a high tunability\cite{Luthi-Acoustic}. The involvement of $F_{\mathbf{m}-\varepsilon}$ causes a response of $\mathbf{m}$ to stress applied. In addition, different magnetic domains, even though they are of nearly degenerate metastable low-energy states at ambient, likely respond differently to stress. These facts make pressure-tuning the landscape of DW possible.

When pressurized, the normal Hall effect of CeAlGe remains essentially unchanged, the magnitude of LTHE reduces to about 0.5 $\mu\Omega\cdot$cm, but what attracts us most is that the LTHE broaden at 1.11 GPa [Fig.~3(d)], and finally it splits into two regions at higher pressure [Figs.~3(e-f)]. At 2.52 GPa, the highest pressure we can reach, the first LTHE is observed between 0 and 0.5 T, while the second one is visible in the range 0.8$\leq$$H$$\leq$1.7 T. To the best of our knowledge, such kind of LTHE evolution with pressure has never been reported in other materials including the structural analog CeAlSi where pressure only weakens the LTHE\cite{Piva-CeAlSiPressure}.

%The role of Weyl nodes to LHE was verified by the sample-dependent measurements which revealed that the LHE is visible only in samples where the Fermi energy lies near the Weyl nodes\cite{Yang-CeAlSiLHE}. We infer that this same scenario also applies to CeAlGe. Note that the presence of DWs in CeAlGe and their coupling to Weyl nodes have been suggested by singual angular magnetoresistance measurements\cite{Suzuki-CeAlGeADMR}.

Before discussing more about the origin of LTHE, we now turn to the resistivity under external magnetic field $\mathbf{H}\parallel \mathbf{c}$, the results of which are also shown in Fig.~3. For example, in the panel (a), we present $\rho_{xx}(T)$ at $p$=0 and different fields. When a small field 0.3 T is applied, the AF transition is suppressed to lower temperature, while another salient feature is that at about 2.5 K, the resistivity drops again. Such a drop becomes more obvious at 0.4 T as shown in the inset to Fig.~3(a). The characteristic temperature for this behavior is denoted by $T_m$. We infer that such a drop in resistivity is associated with field-induced canting of Ce moments, which reduces the spin-flip scattering, and indeed, this is evidenced by the fact that $T_m$ shifts up as field increases. To better describe the evolution of $T_N$ and $T_m$ with field, we show a phase diagram constructed by the false contour plot of $\rho_{xx}(H,T)$ in Fig.~3(c). One clearly finds that $T_N$ and $T_m$ meet at about 0.8 T, above which the Ce moments tend to be polarized prior to undergoing the AF transition, and the signatures for $T_N$ and $T_m$ vanish. The cross-over of field-induced polarization is denoted by $T_P$, which increases with field, as expected. Therefore, (0.8 T, 4.56 K) defines a tricritical point (TCP) where the three phases PM, AF and FIF merge. An clearer determination of $T_N$, $T_m$ and $T_P$ can be given by the $d\rho_{xx}/dT$ plots shown in Fig.~S3. When under pressure, one striking feature is that the AF order seems more robust to field. This can be clearly seen from the data of 2.52 GPa, e.g. under 1.5 T [cf inset to Fig.~3(b)], the Ce moments have been already polarized near 6 K, but they re-enter the AF ordering at $T_N\approx4.9$ K, and this is followed by a canting at $T_m\sim4$ K. Such an AF reentrance broadens the AF region on the phase diagram as shown in Fig.~3(f), and furthermore, the point where $T_N(H)$ and $T_m(H)$ meet [which is denoted by ($H^{\ast},T^{\ast}$) on the phase diagram] now is well separated from the TCP [Fig.3(c-f)]. We note that the field window where the second LTHE shows up coincides with the region spanned by TCP and ($H^{\ast}, T^{\ast}$), and this is so for all the pressures we have measured. For $p$=0, only one LTHE is observed, and the TCP overlaps with ($H^{\ast},T^{\ast}$), indeed.

%Compared with CeAlSi\cite{Yang-CeAlSiLHE,Piva-CeAlSiPressure}, the LHE seen in CeAlGe contains several salient features. (\rmnum{1}) At ambient pressure, the magnitude of LHE in CeAlGe reaches 1.1 $\mu\Omega\cdot$cm at 2 K, about three times larger than in CeAlSi. (\rmnum{2}) The field window for LHE is between 0.3 and 0.8 T for $p$=0, however, under pressure, the onset field of (the 1st) LHE moves towards to $\mu_0H$=0. This is different to CeAlSi where LHE shows up immediately when external field is applied, whenever at ambient or pressurized condition. (\rmnum{3}) Most importantly, in CeAlSi, pressure gradually suppresses the LHE\cite{Piva-CeAlSiPressure}, while in CeAlGe, although the magnitude of LHE is also reduced by pressure, a second LHE region is turned on at higher field which is concomitant with the reentrance of AF phase. According to the discussions here above, the switches on/off of the LHE signals strongly suggest the tunablility of domain-wall chirality by pressure, and this is realized by the magnetoelastic coupling: The pressure-enhanced $T_N$ implies that the AF exchange between Ce moments strengthens. This makes the AF order more competitive for the ground state under pressure, and causes the invasion of AF phase into the field-induced ferromagnetic (FIF) phase, as shown in Fig.~3(f). Chiral domain walls probably appear as the re-entered AF domains nucleate, and are responsible for the LHE. Of course, to further clarify this issue, more investigations based on microscopic techniques are needed.

To gain further understanding of the pressure effect on CeAlGe, we have also performed first-principles calculations using DFT+$U$ method. We investigated 4 competing magnetic structures, which are consistent with the symmetry proposed by the neutron scattering experiment, dubbed AF$^{xy}$, FM$^{xy}$, AF$^z$ and FM$^z$. The AF$^{xy}$ configuration has $\mathbf{m}_A=(m_x, m_y, 0)$ and $\mathbf{m}_B=(-m_y, -m_x, 0)$; FM$^{xy}$ is $\mathbf{m}_A=(m_x, -m_y, 0)$ and $\mathbf{m}_B=(m_y, -m_x, 0)$; AF$^z$ is $\mathbf{m}_A=(0, 0, m_z)$ and $\mathbf{m}_B=(0, 0, -m_z)$; and FM$^z$ has $\mathbf{m}_A=\mathbf{m}_B=(0, 0, m_z)$, where $m_i>0$ ($i=x, y, z$). Among these states, the AF$^{xy}$ state has the lowest energy at both 0 GPa and 3 GPa (TAB. ~1), qualitatively consistent with what neutron scattering experiment suggested\cite{Suzuki-CeAlGeADMR}. We note that the cancellation between $\mathbf{m}_A$ and $\mathbf{m}_B$ is not complete at the ground state, and the actual configuration is ferrimagnetic with residue moment of $\sim0.1\mu_B$. At ambient pressure, the first excited state is FM$^{xy}$, which is only 10.5 meV/f.u. higher than AF$^{xy}$%, consistent with the experimental observation of small AF phase region at ambient pressure.
. At 3 GPa, the first excited state is still FM$^{xy}$, but the state energy is 34.2 meV/f.u. higher than AF$^{xy}$, suggesting the AF$^{xy}$ state is further stabilized under pressure, in line with our experimental observations. We must point out that the noncollinear magnetic energies are extremely subtle in these compounds. However, for RKKY interactions, increased pressure leads to reduced distance between atoms, leading to enhanced hybridization strength and $J_K$ thereafter. Thus, it is generally reasonable to assume an increased magnetic interaction under pressure in local-moment systems, which is consistent with experimental observations. The sensitive magnetic energy with respect to input parameter also suggests that the emergence of domains with different magnetic configurations is highly possible, paving the way for the DW-induced LTHE.

\begin{table}[!hp]
    \centering
    \begin{tabular}{c||c|c|c||c|c|c||c|c||c|c}
      P (GPa) & \multicolumn{3}{c||}{AF$^{xy}$} & \multicolumn{3}{c||}{FM$^{xy}$} & \multicolumn{2}{c||}{AF$^z$} & \multicolumn{2}{c}{FM$^z$} \\
      \hline
       & $E$ & $m_x$ & $m_y$ & $E$  & $m_x$ & $m_y$ & $E$  & $m_z$ & $E$  & $m_z$  \\
      \hline\hline
       0 & 0 & 0.52  & 0.42  & 10.5 & 0.51  & 0.42  & 49.1 & 0.46  & 68.2 & 0.48  \\
       3 & 0 & 0.45  & 0.38  & 34.2 & 0.48  & 0.40  & 42.5 & 0.40  & 44.4 & 0.41  \\
    \end{tabular}
    \caption{Total energy (in meV/f.u.) and magnetic moment of Ce atom (in $\mu_B$) for different configurations. $m_x$, $m_y$ and $m_z$ in each configuration follow the definition in the main text, and contain only spin contribution. }
    \label{tab:magen}
\end{table}

The electronic structure of different magnetic states are also investigated. We assume the most relevant magnetic states in this context are AF$^{xy}$ - the ground state without field - and FM$^z$ when the Ce moments are polarized in the presence of large magnetic field $\mathbf{H}\parallel\mathbf{c}$. The calculated band structure of AF$^{xy}$ at ambient pressure is in good agreement with literature\cite{ChangG-RAlGeDFT} (Fig.~4). At higher pressures (3 GPa), the band structures for each magnetic states are similar to those at ambient pressure. We searched the Brillouin zone (BZ) for possible nodal points in CeAlGe at different magnetic states and pressures. In AF$^{xy}$ state, there are 2 generating operations for the magnetic symmetry, i.e. $\{\mathcal{T}|C_2^z\}$, $\{\sigma_{xy}|\tau\}$, where $\mathcal{T}$ is the time-reversal operator, $C_2^z$ denotes $C_2$ rotation around $z$, $\sigma_{xy}$ is the mirror operation with respect to the $x=y$ plane, and $\tau=(3/4, 1/4, 1/2)$ is the fractional translation. In FM$^{z}$ state, the 2 generating operations are $\{C_4^z|\tau\}$ and $\{\mathcal{T}|\sigma_{x}$\}. Therefore, it is expected that the Weyl node evolution under pressure and field-driven magnetic states involves not only the translation of Weyl points, but also creation/annihilation of a certain nodal points. For example, at 0 GPa AF$^{xy}$ state, we found 20 pairs of Weyl points as confirmed by the chirality calculations [Fig.~4(d)]. At higher pressure up to 3 GPa, the number of topological nodal points retains for the same magnetic structure [Fig.~4(d)]. However, when the system
changes into FM$^z$, the number of Weyl nodes reduces to only 16 pairs, while the other 4 pairs of Weyl points close to the BZ boundary annihilate each other. %Furthermore, the topological Hall coefficient due to Berry curvature was calculated using the Kubo formula for AF$^{xy}$, FM$^{xy}$ and FM$^{z}$ states under ambient pressure and 3 GPa.
%At ambient pressure, $\sigma_{xy}=-0.1$/69.2 $\Omega^{-1}\cdot\mathrm{cm}^{-1}$ for AF$^{xy}$/FM$^{z}$ states.

Returning to the origin of the LTHE. How do DWs facilitate the observed LTHE in CeAlGe? We propose multiple scenarios for this. First of all, the role of scalar spin chirality cannot be disregarded. Itinerant electrons travelling through a landscape of non-coplanar magnetic order acquire a geometrical Berry phase, and can lead to the topological Hall effect independent to the spin-orbit coupling. This scenario has been employed for interpreting the topological Hall effects in many spin systems with non-coplanar ordering, like Nd$_2$Mo$_2$O$_7$\cite{Taguchi-Nd2Mo2O7THE}, SrRuO$_3$ film\cite{Matsuno-SrRuO3THE,Wang-SrRuO3THE},  Mn(Si,Ge)\cite{Lee-MnSiTHE,Neubauer-MnSiTHE,Schulz-MnSiTHE,Kanazawa-MnGeTHE}, UCu$_5$\cite{Ueland-UCu5THE}, Gd$_2$PdSi$_3$\cite{Kurumaji-Gd2PdSi3THE} etc. Compared to these well-known examples, a striking feature of the THE in CeAlGe is the loop shape. This can be explained by the presence of DW whose magnetization is history dependent. As we already discussed above, there are many competing magnetic states in CeAlGe whose energies are close to the AF$^{xy}$ ground state. DWs are prone to form, especially under field. When the external magnetic field is applied, parts of the magnetic moments tend to align in $z$-direction. The coexistence of both AF$^{xy}$ and FM$^z$ domains leads to the DWs. For instance, schematics of such DWs are displayed in the bottom of Figs. 4(e) and (f). Following what was suggested in CeAlSi\cite{Sun-CeAlSiDW}, we also assume the DWs along [100] and [110], respectively. The gradual reorientation of Ce moments across the DWs results in the non-coplanar spin texture with finite chirality $\chi=\mathbf{S_1}\cdot(\mathbf{S_2}\times\mathbf{S_3})$\cite{Ye-Berry,Taguchi-Nd2Mo2O7THE}, which in turn causes the LTHE. When field is strong enough, all the magnetic moments are polarized, the DWs vanish, and the LTHE also disappears. Under pressure, the enhanced AF interaction leads to a more robust AF$^{xy}$ state, and causes the invasion of AF phase into the FIF [which is equivalently deemed as FM$^z$) phase, as shown in Fig.~3(f)]. The region spanned by TCP and ($H^{\ast}$, $T^{\ast}$) delineates the field window where these competing magnetic states coexist \textit{spatially}. Such a reentrance of AF$^{xy}$ invokes new DWs, and this explains the emergence of the second LTHE region.

Band-structure topology may also mediate in the domain wall. In the top of Figs.~4(e) and (f), we show sketches to project the bulk Weyl points in AF$^{xy}$ and FM$^z$ states onto their DWs along [100] and [110], respectively. For clarity, we only show the Weyl points at $k_z$=0, and the distance between them have been exaggerated. Since the Weyl points in AF$^{xy}$ and FM$^z$ states are different in both amounts and locations, they map to the DWs at different positions. The mapped Weyl points from the FM$^z$ states all ``pair-annihilate" on both kinds of DWs, whereas only four Weyl points projected from those close to the BZ boundary in AF$^{xy}$ can survive. Chiral Fermi arcs are expected to connect these Weyl points on the DW, and thus realize a novel confinement of electrons. Such a conducting channel in DW is predicted to cause peculiar physical properties, such as LTHE\cite{Yamaji-MetallicDW}.

For both scenarios, the presence of chiral domain walls plays a key role for the LTHE in CeAlGe. We must admit that it is difficult to tell which is the dominant only by transport measurements. To further clarify this issue, additional experiments based on microscopic imaging and dispersion spectroscopy techniques are both required in the future. However, the observed pressure controllable giant topological Hall effect strongly suggest the capability of switching on/off domain-wall chirality in magnetic Weyl semimetals through magnetoelastic coupling, possessing immense potential in device applications. \\

\noindent
\textbf{4 Conclusion}\\

\noindent
On the prototype of rare-earth-contained magnetic Weyl semimetal CeAlGe, we studied the transport and thermodynamic properties under pressure. Large loop-shaped topological Hall effect was observed in CeAlGe, and splits into well separated regions under pressure. The similarity of our CeAlGe and previously reported Nd$_2$Ir$_2$O$_7$ and CeAlSi indicates that such loop-shaped topological Hall effect is a common feature of magnetic Weyl semimetals as a consequence of chiral domain walls. The switches on/off of such unusual electromagnetic response strongly suggest the tunablility of domain-wall chirality in magnetic Weyl semimetals by magnetoelastic coupling, and hopefully this would find applications in electronic/spintronic devices. \\

\noindent
\textbf{Acknowledgments}\\
The authors thank Joe D. Thompson for insightful discussions. This work was supported by the open research fund of Songshan Lake Materials Laboratory (2022SLABFN27), NSF of China (12274364, U1932155), the Fundamental Research Funds for the Central Universities of China (2019kfyXMBZ071), National Key R\&D Program of China (2022YFA1602602), Guangdong Basic and Applied Basic Research Foundation (2022B1515120020), and the Pioneer and Leading Goose R\&D Program of Zhejiang (2022SDXHDX0005).

\emph{}\\

\noindent
%\textbf{Data availability}\\
%The data that support the plots within this paper and other findings of this study are available from the corresponding authors upon reasonable request.

\normalem
%\bibliography{biblio.bib}

\noindent
%\textbf{Author contributions}\\
%Y. Luo conceived the project and designed the experiments. X. He and Y. Li synthesized the material. X. He performed most of the measurements. C. Cao carried out first-principles calculations. H. Zeng and Z. Zhu provided constructive suggestions. X. He, C. Cao and Y. Luo analyzed the data, discussed the results and wrote the paper with input from all the authors.

\noindent
%\textbf{Author information}\\
%The authors declare no competing financial interests. Correspondence and requests for materials should be addressed to C. Cao (ccao@zju.edu.cn) and Y. Luo (mpzslyk@gmail.com).

\newpage
\textbf{Figures:}

\begin{figure}[!h]
\hspace*{-10pt}
\includegraphics[width=12cm]{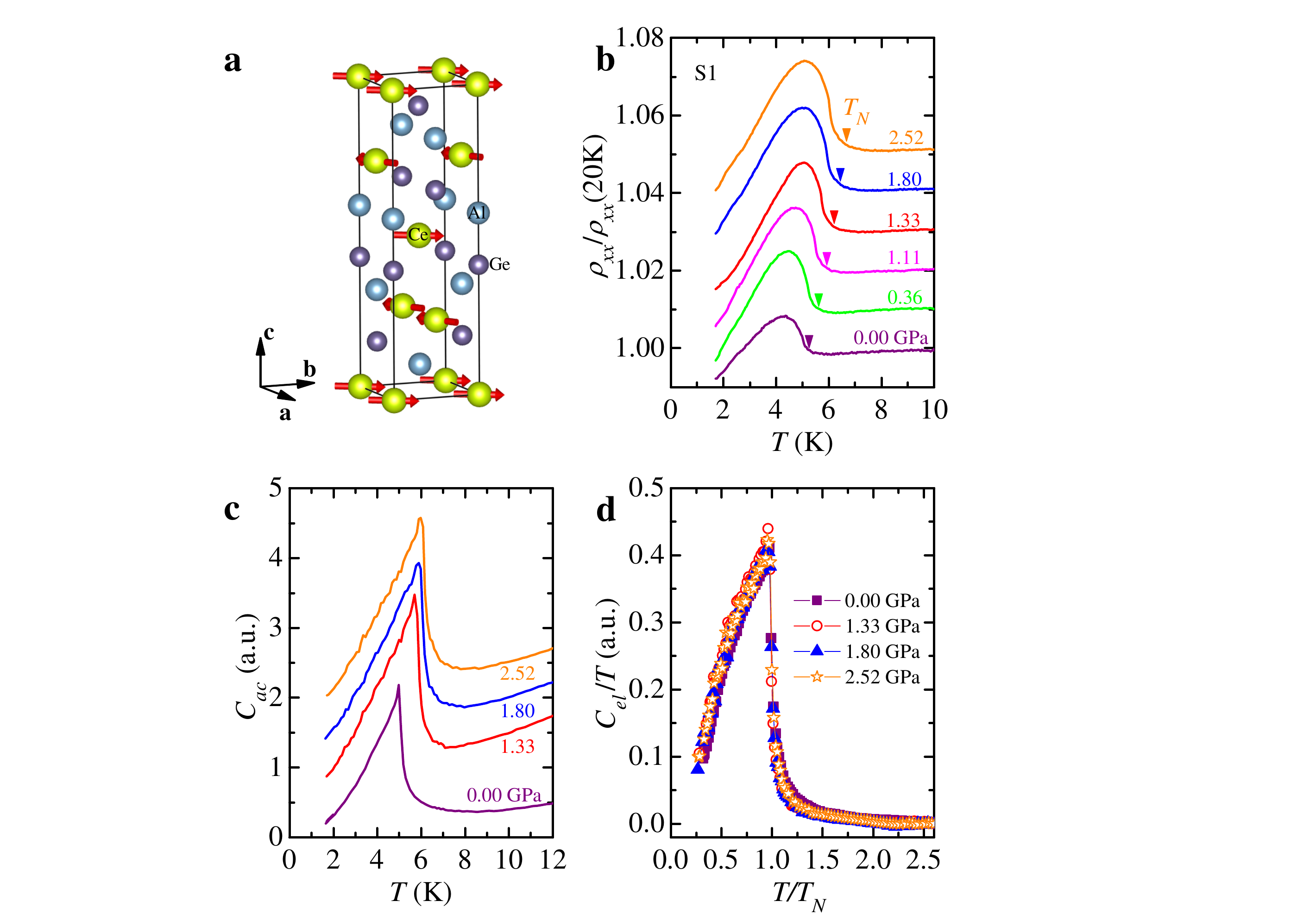}
\label{Fig1}
\end{figure}
\vspace*{0pt}
\textbf{Figure 1} (Color online) Pressure effect of the magnetic transition in CeAlGe. (a) Schematic magnetic structure of CeAlGe. The order parameters of the two Ce sublattices are $\mathbf{m_A}=(m_x,m_y,0)$ and $\mathbf{m_B}=(-m_y,-m_x,0)$, of the coplanar magnetic structure wih $Fd'd2'$ magnetic space group\cite{Suzuki-CeAlGeADMR}. (b) and (c) display temperature dependent normalized resistivity $\rho_{xx}/\rho_{xx}(20\text{K})$ and AC heat capacity $C_{ac}$ under various pressures. The arrows in (b) mark the AF transition at $T_N$. The curves have been vertically shifted for clarity. (d) Scaling plot of $C_{el}/T$ vs $T/T_{N}$ showing that the curves for different pressures collapse into a single line.  \\

\newpage

\begin{figure}[!h]
\hspace*{-10pt}
\includegraphics[width=17cm]{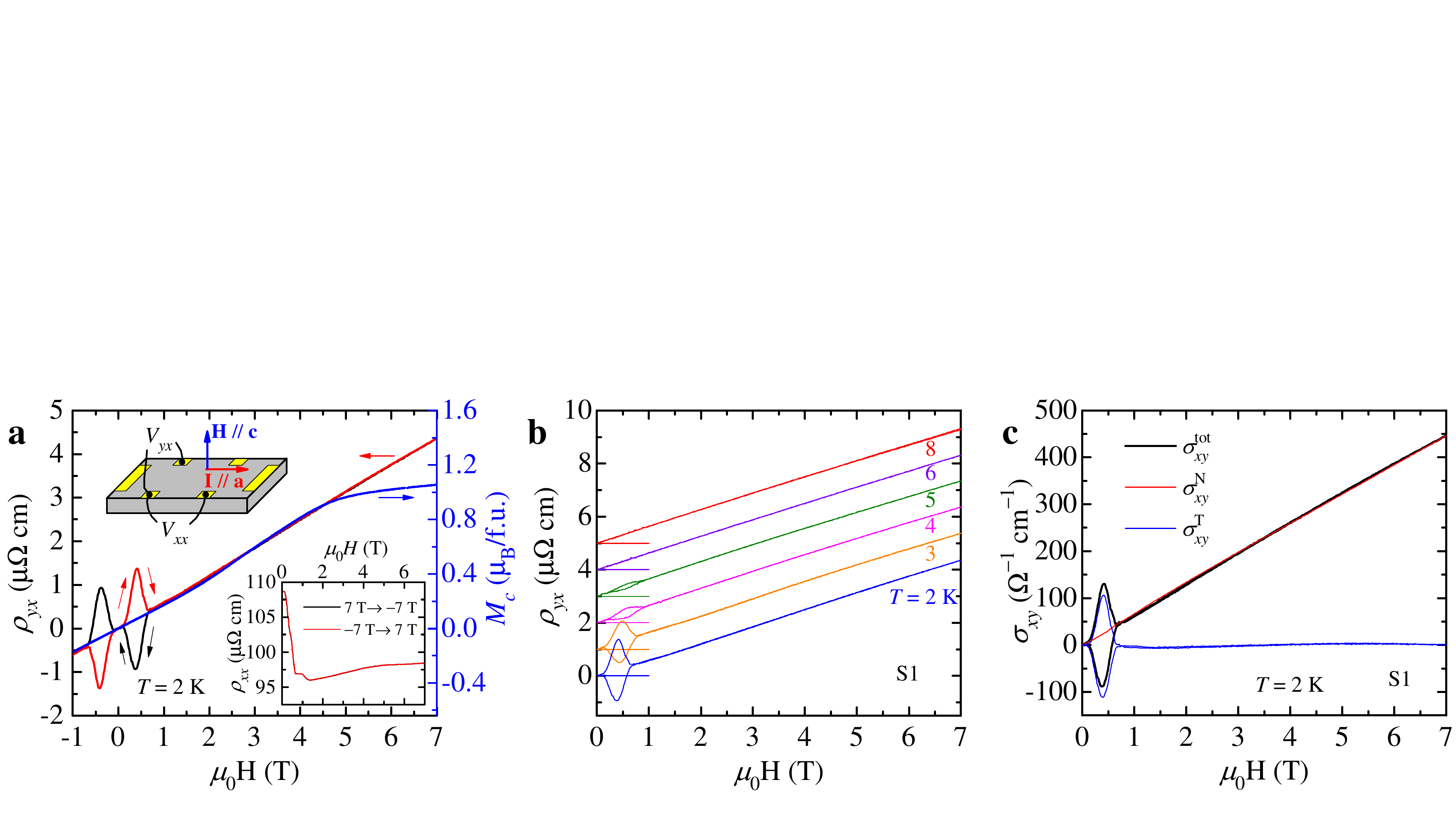}
\label{Fig2}
\end{figure}
\vspace*{-20pt}
\textbf{Figure 2} (Color online) Magneto-transport properties of CeAlGe at atmosphere. (a) Isothermal $T$=2 K field dependent Hall resistivity $\rho_{yx}$ (left axis) and magnetization $M_c$ (right axis). Note that a hysteresis loop is seen between 0.15 and 0.7 T in $\rho_{yx}(H)$, which is absent in $M_c(H)$. The left inset shows the configuration of the measurements. The bottom inset shows $\rho_{xx}$ as a function of field. (b) Hall effect at different temperatures. For clarity, the curves have been shifted vertically. (c) Hall conductivity $\sigma_{xy}$ vs. $H$ at 2 K, where $\sigma_{xy}=\rho_{yx}/(\rho_{xx}^2+\rho_{yx}^2)$. The total Hall conductivity $\sigma_{xy}^{tot}$ decomposes into normal Hall conductivity $\sigma_{xy}^{N}$ and loop-shaped topological Hall conductivity $\sigma_{xy}^{T}$. \\

\newpage

\begin{figure}[!h]
\hspace*{-20pt}
\includegraphics[width=17cm]{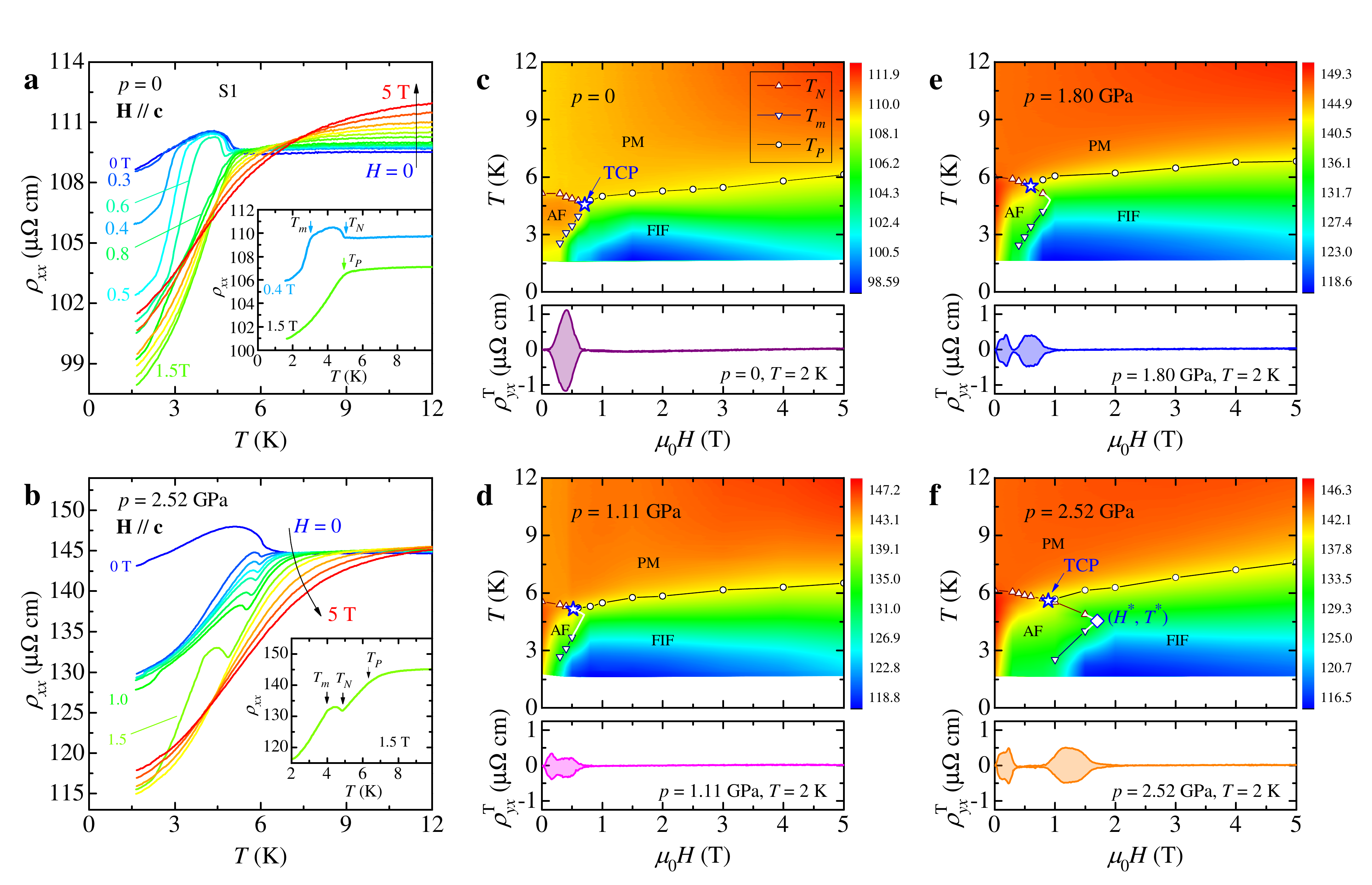}
\label{Fig3}
\end{figure}
\vspace*{-20pt}
\textbf{Figure 3} (Color online) Magneto-transport properties of CeAlGe under pressure. (a-b) $\rho_{xx}(T)$ measured at different magnetic fields for $p$=0 and 2.52 GPa, respectively. The inset of (a) shows the definition of $T_N$, $T_m$ and $T_P$. The top panels of (c-f) display the false contour plots of $\rho_{xx}(H, T)$ for $p$=0, 1.11, 1.80, and 2.52 GPa. The abbreviations are: AF = antiferromagnetic, PM = paramagnetic, FIF = field-induced ferromagnetic, and TCP = tri-critical point. The regions spanned by TCP and ($H^{\ast}$, $T^{\ast}$) are the field window where competing magnetic states AF and FIF \textit{spatially} coexist. The bottom panels are the field dependence of $\rho_{yx}^{T}$ at 2 K. \\

\newpage

\begin{figure}[!h]
\hspace*{-20pt}
\includegraphics[width=17cm]{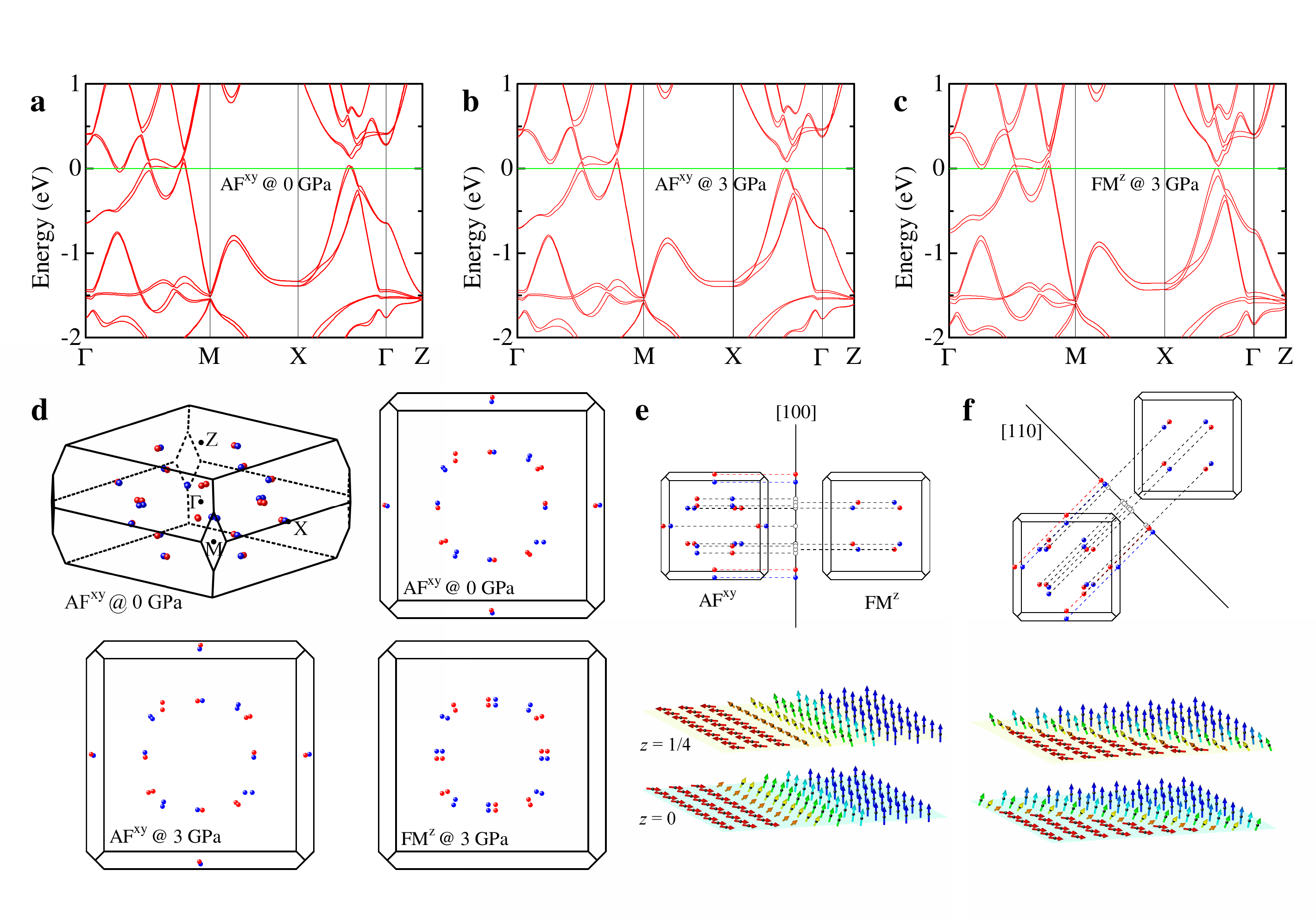}
\label{Fig4}
\end{figure}
\vspace*{-20pt}
\textbf{Figure 4} (Color online) DFT calculations of CeAlGe at 0 and 3 GPa. Panels (a-c) show band structure in AF$^{xy}$ at 0 GPa, AF$^{xy}$ at 3 GPa, and FM$^{z}$ at 3 GPa, respectively. The calculations were performed with spin-orbit coupling. (d) Location of Weyl points for conditions at (a-c). (e) Sketch of spin texture for domain wall along [100], and the projections of Weyl points on the domain wall. The red and blue circles characterize different chiralities, and the black circles are non-chiral. For clarity, only the Weyl points at $k_z=0$ are shown, and the interval between them are exaggerated. (f) ibid, but for domain wall along [110].\\

\newpage

\setcounter{table}{0}
\setcounter{figure}{0}
\setcounter{equation}{0}
\setcounter{section}{0}
\renewcommand{\thefigure}{S\arabic{figure}}
\renewcommand{\thetable}{S\arabic{table}}
\renewcommand{\theequation}{S\arabic{equation}}
\onecolumngrid

\begin{center}
{\it\textbf{Supplementary Information: }}
\textbf{Pressure-tuning domain-wall chirality in noncentrosymmetric magnetic Weyl semimetal}\\

\end{center}

\begin{center}

Xiaobo He$^{1}$, Yuke Li$^{2}$, Hai Zeng$^{1}$, Zengwei Zhu$^{1}$, Shiyong Tan$^{3}$, Yongjun Zhang$^{4}$, \\ Chao Cao$^{5*}$,Yongkang Luo$^{1\dag}$\\
\footnotesize{$^1${\it Wuhan National High Magnetic Field Center and School of Physics, Huazhong University of Science and Technology, Wuhan 430074, China;}}\\
\footnotesize{$^2${\it School of Physics and Hangzhou Key Laboratory of Quantum Matter, Hangzhou Normal University, Hangzhou 311121, China;}}\\
\footnotesize{$^3${\it Science and Technology on Surface Physics and Chemistry Laboratory, Mianyang 621908, China;}}\\
\footnotesize{$^4${\it Institute for Advanced Materials, Hubei Normal University, Huangshi 435002, China; and}}\\
\footnotesize{$^5${\it Center for Correlated Matter and School of Physics, Zhejiang University, Hangzhou 310058, China.}}

%\date{\today}
\end{center}

%\textbf{SI \Rmnum{1}: S\lowercase{ample characterization}}

\vspace*{-37pt}
\begin{figure}[!h]
\hspace*{-15pt}
\vspace*{-20pt}
\includegraphics[width=14.0cm]{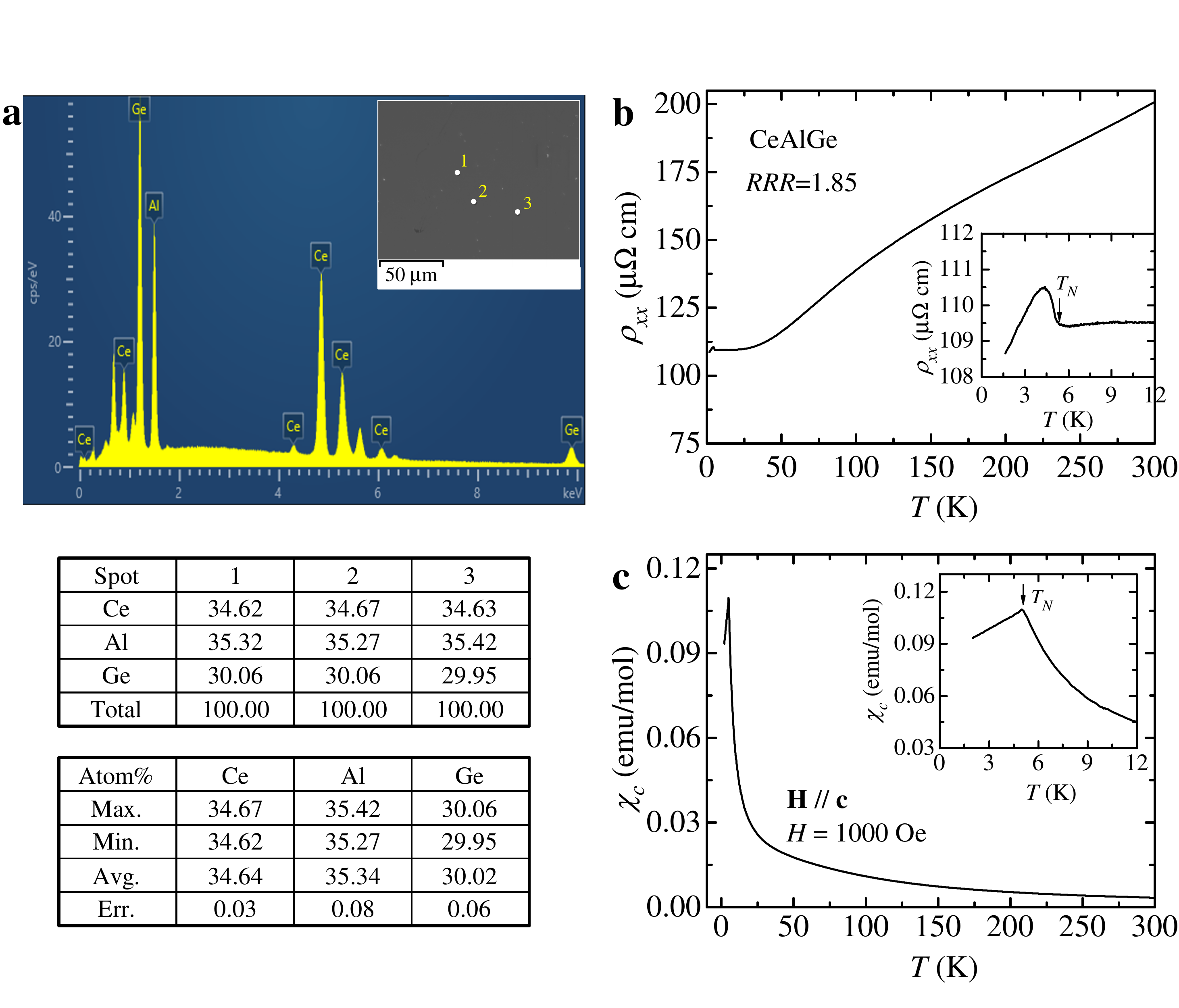}
\label{FigS1}
\end{figure}
\vspace*{-00pt}
\textbf{Figure S1} (Color online) Sample characterization. (a) EDS results of a selected CeAlGe single crystal. The atomic concentrations of Ce, Al, and Ge are shown in the bottom tables. (b) Temperature dependence of in-plane resistivity $\rho_{xx}$. The inset shows a zoom-in plot for the low-temperature AF transition. (c) Magnetic susceptibility $\chi_c$ as a function of $T$. A peak is visible at $T_N$=5.1 K, characteristic of AF transition.  \\

%\textbf{SI \Rmnum{3}: D\lowercase{efinition of critical temperature}}
\newpage

\vspace*{-40pt}
\begin{figure}[!h]
\hspace*{-10pt}
\vspace*{-15pt}
\includegraphics[width=16cm]{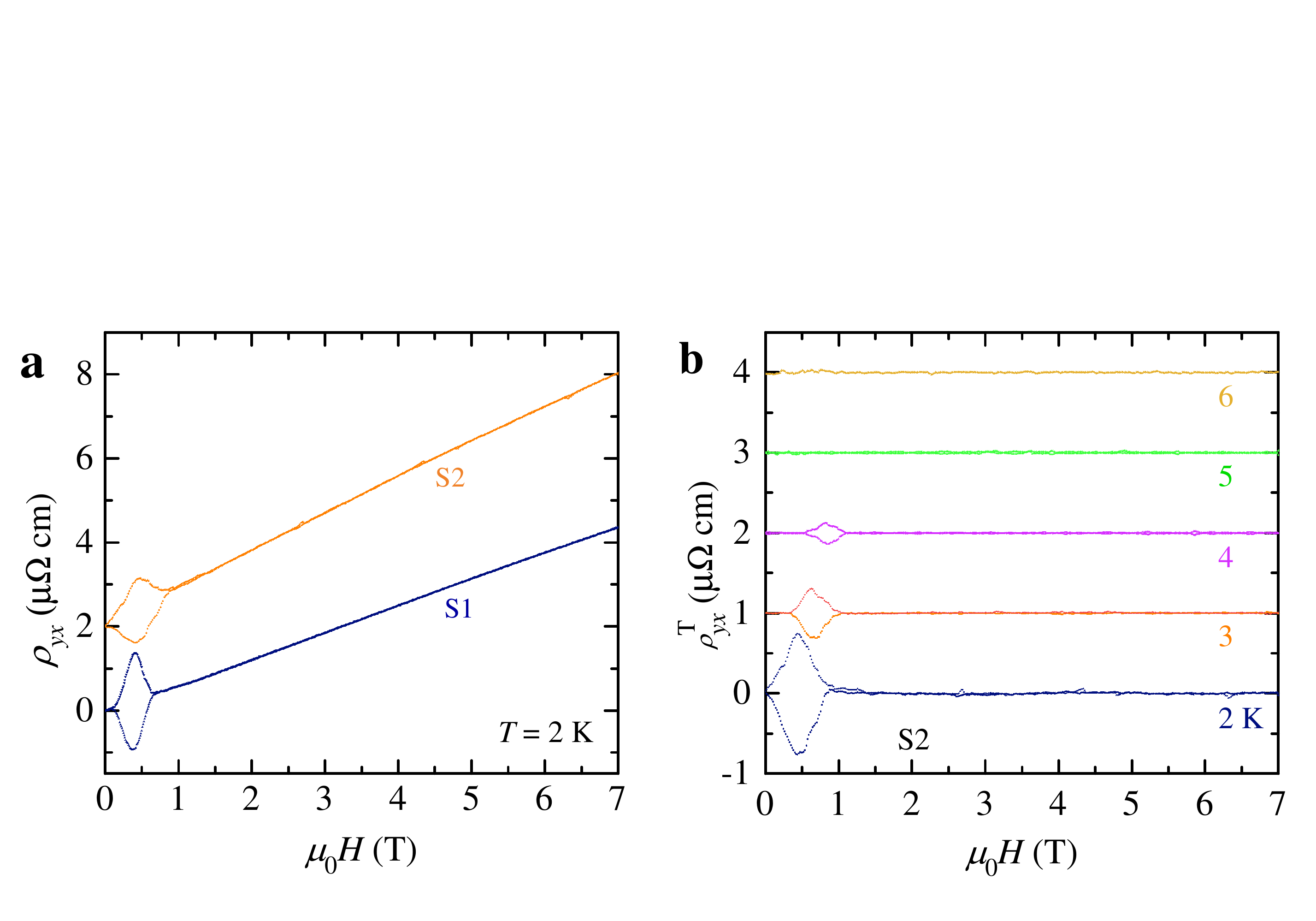}
\label{FigS2}
\end{figure}
\vspace*{-0pt}
\textbf{Figure S2} (Color online) LTHE observed on different CeAlGe samples. (a) Hall resistivity at 2 K measured on samples S1 and S2. They both show loop-shaped topological Hall effect, but the size and the shape of the loop are sample-dependent. (b) Field dependence of $\rho_{yx}^T$ of sample S2 at different temperatures. The position of the loop moves to higher field as $T$ raises, qualitatively consistent with the evolution of the AF-FIF boundary.\\

\newpage

\vspace*{-20pt}
\begin{figure}[!h]
\hspace*{-10pt}
\vspace*{-15pt}
\includegraphics[width=16cm]{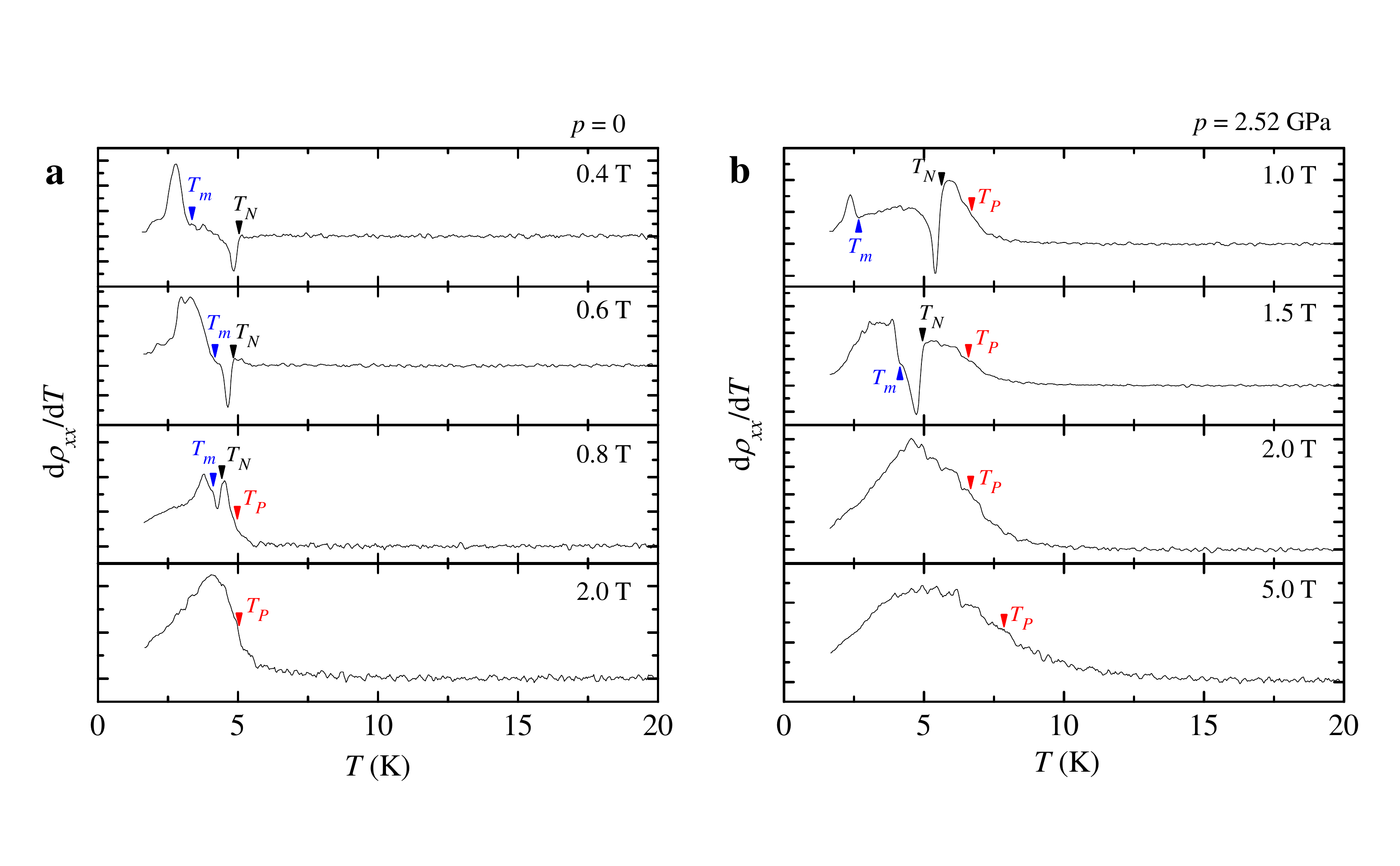}
\label{FigS3}
\end{figure}
\vspace*{-10pt}
\textbf{Figure S3} (Color online) Definition of the critical temperatures. $d\rho_{xx}/dT$ curves for representative field strengths. The left column is for $p$=0, and the right column is for $p$=2.52 GPa. $T_N$ is defined at the position where $d\rho_{xx}/dT$ starts to drop; $T_m$ is defined at the inflection point in $d\rho_{xx}/dT$; $T_P$ is defined near the mid-point of the upturn in $d\rho_{xx}/dT$. \\

%\newpage

\end{document}